\documentclass[12pt]{article}
\usepackage{amsmath,amssymb}
\usepackage{graphicx}
\renewcommand{\thefootnote}{\fnsymbol{footnote}}
\newcommand{\bea}{\begin{eqnarray}}
\newcommand{\ena}{\end{eqnarray}}
\newcommand{\vs}[1]{\vspace{#1 mm}}

\newcommand{\PL}[1]{Phys.\ Lett.\ {\bf #1}}

\newcommand{\PR}[1]{Phys.\ Rev.\ {\bf #1}}

\begin{document}
\noindent

\begin{titlepage}
\setcounter{page}{0}
\begin{flushright}
September,  2001\\
hep-ph/0109244\\
OU-HET 396\\
\end{flushright}
\vs{2}
\begin{center}
{\Large{\bf $\bar B \to D \tau \bar\nu_{\tau}$ in two-Higgs-doublet models}}
\footnote{Talk given by T. Miura at 
Workshop on Higher Luminosity B Factory, 
 August 23-24, 2001, KEK, Japan}\\
\vs{6}
{\large
Takahiro MIURA\footnote{e-mail address:
miura@het.phys.sci.osaka-u.ac.jp}
and Minoru TANAKA\footnote{e-mail address:
tanaka@phys.sci.osaka-u.ac.jp}\\
\vs{2}
{\sl Department of Physics,
Osaka University \\ Toyonaka, Osaka 560-0043, Japan}
}
\end{center}
\vs{8}
%\centerline{{\bf Abstract}}  
\begin{abstract}
We study the exclusive semi-tauonic $B$ decay, 
$\bar B \to D \tau \bar\nu_{\tau}$, in two-Higgs-doublet models.
Using recent experimental and theoretical results on 
hadronic form factors, 
we estimate theoretical uncertainties in the branching ratio.
As a result, we clarify the potential sensitivity of this mode 
to the charged Higgs exchange.
Our analysis will help to probe the charged Higgs boson 
at present and future B factory experiments.
\end{abstract}

\end{titlepage}
\newpage
\vskip 2cm
\renewcommand{\thefootnote}{\arabic{footnote}}
\setcounter{footnote}{0}

\section{Introduction}
Many interesting models for the new physics 
beyond the standard model (SM) have been considered.
One of the most attractive models is
the minimal supersymmetric standard model (MSSM) \cite{MSSM}. 
In the MSSM, two Higgs doublets are introduced in order to 
cancel the anomaly and to give the fermions masses.
The introduction of the second Higgs doublet inevitably means that 
a charged Higgs boson is in the physical spectra.
So, it is very important to study effects of the charged Higgs boson.

Here, we study effects of the charged Higgs boson 
on the exclusive 
semi-tauonic $B$ decay, $\bar B \to D \tau \bar\nu_{\tau}$,
in the MSSM.
In a two-Higgs-doublet model, we have a pair of charged Higgs bosons,
$H^{\pm}$, 
and its couplings to quarks and leptons are given by 
\bea
   {\cal L}_H&=&(2\sqrt{2}G_F)^{1/2}
    \bigl[X\overline{u}_L V_{KM}M_d d_R+
          Y\overline{u}_R M_u V_{KM}d_L
          +Z\overline{\nu}_L M_l l_R\bigr]\;H^+
            \nonumber\\
&&+\mbox{\large h.c.}\;,\label{Lag}
\ena
where $M_u$, $M_d$ and $M_l$ are diagonal quark and lepton mass matrices,
and $V_{KM}$ is Kobayashi-Maskawa matrix \cite{CKM}.
In the MSSM, we obtain 
\bea
   X=Z=\tan\beta\;,\;\;Y=\cot\beta\;,
\ena
where $\tan\beta=v_2/v_1$ is the ratio of the vacuum expectation 
values of the Higgs bosons. Since the Yukawa couplings of the MSSM are 
the same as those of the so-called Model II of 
two-Higgs-doublet models \cite{HHG}, 
the above equations and the following results apply to the latter as well.

From these couplings, we observe that 
the amplitude of charged Higgs exchange 
in $\bar B \to D \tau \bar\nu_{\tau}$ has a term proportional to 
$m_b \tan^2 \beta$. Thus, the effect of the charged Higgs boson is 
more significant for larger $\tan\beta$. 

In Sec.2, we give formula of the decay rate. 
The employed hadronic form factors are described in Sec.3.
In Sec.4, we show our numerical results. Sec.5 is devoted to conclusion.

\section{Formula of the decay rate}

Using the above Lagrangian in Eq.(\ref{Lag}) and 
the standard charged current Lagrangian, 
we can calculate the amplitudes of charged Higgs exchange and 
$W$ boson exchange in $\bar B \to D \tau \bar\nu_{\tau}$.

The $W$ boson exchange amplitude is given by \cite{Hagiwara}
\bea
   {\cal M}_{s}^{\lambda_\tau}(q^2,x)_W=
   \frac{G_F}{\sqrt{2}}V_{cb} 
   \sum_{\lambda_W}\eta_{\lambda_W} 
   L_{\lambda_W}^{\lambda_\tau}H_{\lambda_W}^{s}\;,\label{amp1}
\ena
where $q^2$ is the invariant mass squared of the leptonic system, and
$x=p_B\cdot p_\tau/m_B^2$.
The $\tau$ helicity and the virtual $W$ helicity are denoted by 
$\lambda_\tau=\pm$ and $\lambda_W=\pm ,\,0,\,s$, and 
the metric factor $\eta_{\lambda_W}$ is given by 
$\eta_{\pm}=\eta_0=-\eta_s=1$.
The hadronic amplitude which describes $\bar B\rightarrow D\,W^*$ 
and the leptonic amplitude which describes 
$W^*\rightarrow\tau\bar\nu$ are given by
\bea
   H_{\lambda_W}^{s}(q^2)&\equiv&
   \epsilon_\mu^*(\lambda_W)\langle D(p_D)|
                    \bar c\gamma^\mu b|\bar B(p_B)\rangle\;,\label{hadamp}\\
   L_{\lambda_W}^{\lambda_\tau}(q^2,x)&\equiv&
   \epsilon_\mu(\lambda_W)\langle\tau(p_\tau,\lambda_\tau)\bar\nu_\tau(p_\nu)|
               \bar \tau\gamma^\mu(1-\gamma_5)\nu_\tau|0\rangle\;,
\ena
where $\epsilon_\mu(\lambda_W)$ is the polarization vector of the virtual
$W$ boson.

The charged Higgs exchange amplitude is given by \cite{Tanaka}
\bea
  {\cal M}_{s}^{\lambda_\tau}(q^2,x)_H=
  \frac{G_F}{\sqrt{2}}V_{cb} L^{\lambda_\tau} 
  \Bigl[X Z^*\frac{m_b m_\tau}{M_{H}^2}
                 H_R^{s}+
                 Y Z^*\frac{m_c m_\tau}{M_{H}^2}
                 H_L^{s}\Bigr]\;.\label{amp2}
\ena
Here, the hadronic and leptonic amplitudes are defined by 
\bea
   H_{R,L}^{s}(q^2)&\equiv&
   \langle D(p_D)|\bar c(1\pm\gamma_5)b|\bar B(p_B)\rangle\;,\\
  L^{\lambda_\tau}(q^2,x)&\equiv&
   \langle\tau(p_\tau,\lambda_\tau)\bar\nu_\tau(p_\nu)|
   \bar \tau (1-\gamma_5)\nu_\tau|0\rangle\;.
\ena
These amplitudes are related to the $W$ exchange amplitudes as 
\bea
   H_{R,L}^s=\frac{\sqrt{q^2}}{m_b-m_c} H_s^s\;,\;\;
   L^{\lambda_\tau}=\frac{\sqrt{q^2}}{m_\tau}L^{\lambda_\tau}_s\;,
\ena
where the former relation is valid in the heavy quark limit.

Using the amplitudes of Eqs.(\ref{amp1}) and (\ref{amp2}), 
the differential decay rate is given by 
\bea
   \frac{d\Gamma}{dq^2}&=&
   \frac{G_F^2 |V_{cb}|^2 v^4\sqrt{Q_+ Q_-}}{128\pi^3m_B^3}
   \Bigl[\Bigl(\frac23 q^2+\frac13 m_{\tau}^2 \Bigr)(H^s_0)^2 \Bigr.\nonumber\\ 
\Bigl.
&&+m_{\tau}^2\Bigl(\frac{q^2 \tan^2\beta}{M_H^2}\frac{m_b}{m_b-m_c}+\frac{q^2}{M_H^2}\frac{m_c}{m_b-m_c}-1 \Bigr)^2(H_s^s)^2  \Bigr]\;,\label{decayrate}
\ena
where $Q_{\pm}=(m_B\pm m_D)^2-q^2$ and $v=\sqrt{1-m_{\tau}^2/q^2}$.
Note that if $\tan \beta \gtrsim 1$, in which we are interested, 
this decay rate is practically a function of $\tan \beta/M_H$
because the second term in the coefficient of 
$(H^s_s)^2$ is negligible for $m_b\tan^2 \beta \gg m_c$.

\section{Hadronic form factors}

In order to obtain the decay rate numerically, it is necessary to 
calculate the hadronic amplitude in Eq.(\ref{hadamp}). This amplitude is given 
in terms of hadronic form factors:
\bea
\langle D(p_D)|\bar c\gamma^\mu b|\bar B(p_B)\rangle 
   = \sqrt{m_B m_D}\bigl[h_{+}(y)(v+v')^{\mu}+h_{-}(y)(v-v')^{\mu}\bigr]
\;,
\ena
where $v=p_B/m_B$, $v'=p_D/m_D$ and 
$y\equiv v\cdot v'=(m_B^2+m_D^2-q^2)/2 m_B m_D$ \cite{HFF}. 

In the heavy quark limit and in the leading logarithmic approximation, 
$h_{+}(y)$ and $h_{-}(y)$ are given as
\bea
 h_{+}(y)= \xi(y)\;,\;\;h_{-}(y)= 0\;,
\ena
where $\xi(y)$ is the universal form factor \cite{Isgur}.

The form of $\xi(y)$ is constrained strongly by the dispersion relations
as \cite{Caprini}
\bea
 \xi(y)\simeq 1-8\rho_1^2z+(51.\rho_1^2-10.)z^2-(252.\rho_1^2-84.)z^3\;,
\ena
where $z=(\sqrt{y+1}-\sqrt{2})/(\sqrt{y+1}+\sqrt{2})$. 
To determine the slope parameter $\rho_1^2$, 
we use the experimental data of Belle \cite{SPBELLE}, and 
we obtain 
\bea
\rho_1^2=1.33\pm 0.22\;.\label{SP}
\ena
This error of $\rho_1^2$ dominantly 
contributes to the uncertainty in the theoretical calculation of 
the branching ratio.

\section{Numerical results}

\begin{figure}[t]
 \begin{center}
\includegraphics[width=7.8cm]{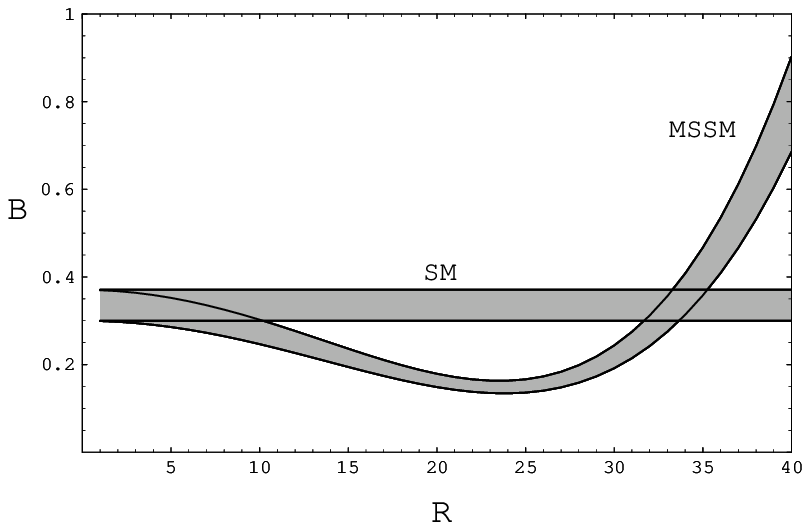} \hspace{0.1cm}
\includegraphics[width=7.8cm]{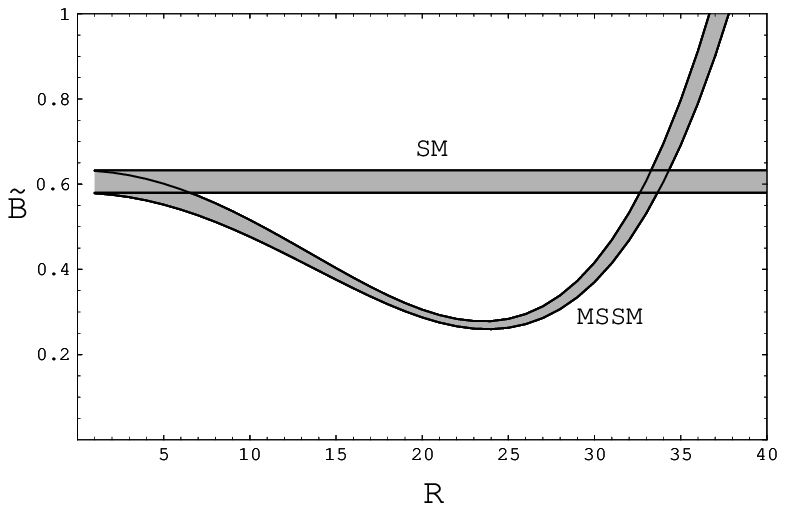}
\caption{The ratios $B$ and $\tilde B$ as functions of $R$ in the MSSM 
  and the SM. The shaded regions show the predictions with 
  the error in the slope parameter $\rho_1^2$ in Eq.(\ref{SP}).
  The flat bands show the SM predictions.
  (a) $B$: the decay rate normalized to 
  $\Gamma(\bar B\rightarrow D\mu\bar\nu_\mu)_{SM}$. 
  (b) $\tilde B$: the same as (a) except that 
  the denominator is integrated over 
  $m_{\tau}^2\leq q^2\leq (m_B-m_D)^2$.}
 \end{center}
\end{figure}

Now, we consider the following ratio, 
\bea
   B=\frac{\Gamma(\bar B\rightarrow D\tau\bar\nu_\tau)}{\Gamma(\bar B\rightarrow D\mu\bar\nu_\mu)_{SM}}\;,\label{Br}
\ena
where the denominator is the decay rate of
$\bar B\rightarrow D\mu\bar\nu_\mu$ in the SM,
since the uncertainties due to 
the form factors and other parameters
tend to reduce or vanish by taking the ratio.

Fig.1(a) is the plot of our predictions of the ratio in Eq.(\ref{Br})
as a function of $R$, which is defined by $R\equiv m_W\tan\beta/m_H$. 
The shaded regions show the MSSM and SM predictions with 
the error in the slope parameter $\rho_1^2$ in Eq.(\ref{SP}).
As seen in Fig.1(a), when $R$ reaches about 32, 
the branching ratio in the 
MSSM becomes the same as the one in the SM. 
It is because the interference of 
the W exchange and the charged Higgs exchange is negative. From Fig.1(a), 
we expect that the experimentally possible sensitivity 
of $R$ is $\sim 10$, provided that the error in $\rho_1^2$ 
will not change.

In Fig.1(b), we also show the ratio, 
\bea
   \tilde B=\frac{\Gamma(\bar B\rightarrow D\tau\bar\nu_\tau)}{\tilde\Gamma(\bar B\rightarrow D\mu\bar\nu_\mu)_{SM}}\;,\label{Br2}
\ena
the same as Fig.1(a), but its denominator is 
$\tilde\Gamma(\bar B\rightarrow D\mu\bar\nu_\mu)_{SM}$,
which is integrated over the same $q^2$ region as the $\tau$ mode, 
i.e., $m_{\tau}^2\leq q^2\leq (m_B-m_D)^2$. From Fig.1(b), we expect 
less theoretical uncertainty and a better sensitivity of $\tilde B$ 
compared with $B$ in Fig.1(a).

\begin{figure}[t]
 \begin{center}
\includegraphics[width=7.8cm]{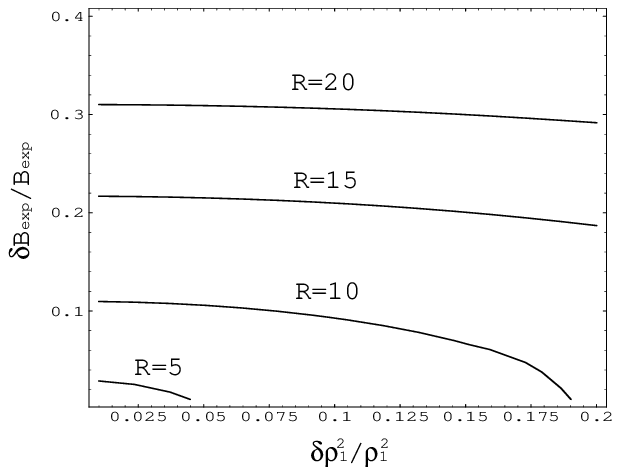} \hspace{0.1cm}
\includegraphics[width=7.8cm]{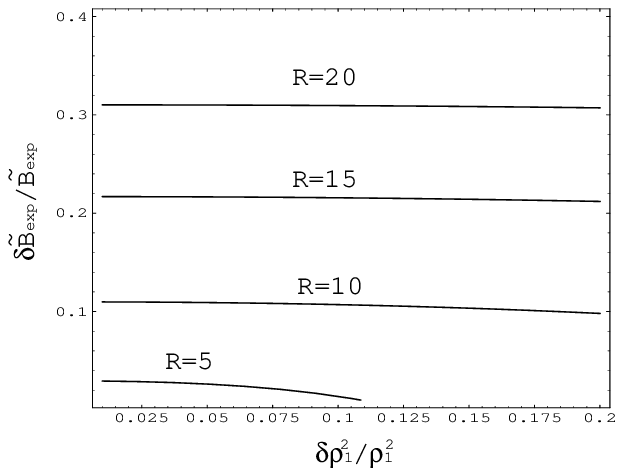}
\caption{(a) The contour plot of upper bound of $R$ at $90\%$ CL 
  as a function of $\delta\rho_1^2$ and $\delta B_{exp}$. 
  (b) The same as (a)
  except that the ratio in Eq.(\ref{Br2}) is used.}
 \end{center}
\end{figure}

Once the experimental values of $B$ ($\tilde B$), its error $\delta B$ 
($\delta \tilde B$), $\rho_1^2$ and its error $\delta\rho_1^2$ are given,
we can obtain a bound on $R$.
In the following, 
we assume the SM prediction as the experimental value of $B$ ($\tilde B$)
, i.e., $B_{exp}=B_{SM}\pm \delta B_{exp}$ 
($\tilde B_{exp}=\tilde B_{SM}\pm \delta \tilde B_{exp}$), and we use
the central value of Eq.(\ref{SP}) as the input of the slope parameter.

Fig.2(a) is the contour plot of upper bound of $R$ at $90\%$ CL 
as a function of $\delta\rho_1^2$ and $\delta B_{exp}$.
From this figure,
if $\delta B_{exp}=0$, and $\delta\rho_1^2 \sim 17\%$, 
which corresponds to the present experimental error in Eq.(\ref{SP}), 
we expect that an upper bound of $R\sim 10$, 
which is consistent with the result of Fig.1(a).
If we will observe $\bar B\rightarrow D\tau\bar\nu_\tau$ 
with $\delta B_{exp}\simeq 20\%$, 
we expect that an upper bound of $R\sim 15$ 
weakly depending on $\delta\rho_1^2$.

In Fig.2(b), we also show a similar contour plot where we use 
the ratio defined in Eq.(\ref{Br2}), i.e., 
normalized to $\tilde\Gamma(\bar B\rightarrow D\mu\bar\nu_\mu)_{SM}$.
We observe that the upper bound of R is almost
independent of $\delta\rho_1^2$ in this case.
Thus, it is important to make the experimental error 
in $B$ ($\tilde B$), 
$\delta B_{exp}$ ($\delta \tilde B_{exp}$), small rather than 
$\delta\rho_1^2$.

\section{Conclusion}
As seen in our numerical results, the branching ratio of
$\bar B\rightarrow D\tau\bar\nu_\tau$
is a sensitive probe of the MSSM-like Higgs sector. 
We expect an upper bound of $R\lesssim 15$ 
when $\delta B_{exp}\simeq 20\%$ is achieved.
So, if $\bar B\rightarrow D\tau\bar\nu_\tau$ is observed 
at a B factory experiment, a significant regions 
of the parameter space of the MSSM Higgs sector will be covered. 
Comparing with the Higgs search scenario of LHC \cite{LHC},
we conclude that present and future B factories are potentially 
competitive with LHC.

As future improvements of the present work,
the $q^2$ distribution \cite{Soni} 
and the $\tau$ polarization \cite{Tanaka}
of $\bar B\rightarrow D\tau\bar\nu_\tau$ are promising.
In these quantities, we will expect that 
the theoretical uncertainties from the error in the 
slope parameter become very small.
However, we should take 1/m and QCD corrections into account.
These corrections are neglected in the present work because 
they lead to smaller uncertainties than those from $\delta \rho_1^2$.
For the $q^2$ distribution and the $\tau$ polarization, they are 
expected to be dominant uncertainties in the theoretical calculations.
These issues will be addressed elsewhere.


\begin{thebibliography}{99}

%%% MSSM
\bibitem{MSSM}
For a review, see, e.g., 
H.~E.~Haber and G.~L.~Kane, Phys. Rep. 117 (1985) 75.

%%% CKM
\bibitem{CKM}
M.~Kobayashi and T.~Maskawa, Prog. Theor. Phys. 49 (1973) 652.

%%% Higgs Hunter's Guide
\bibitem{HHG}
J.~F.~Gunion, H.~E.~Haber, G.~L.~Kane and S.~Dawson, 
{\em The Higgs Hunter's Guide}
(Addison-Wesley Publishing Company, 1990).

%%% calculation of amplitude
\bibitem{Hagiwara}
K.~Hagiwara, A.~D.~Martin and M.~F.~Wade, Nucl. Phys. B327 (1989) 569;\\
K.~Hagiwara, A.~D.~Martin and M.~F.~Wade, Z. Phys. C46 (1990) 299.

\bibitem{Tanaka}
M.~Tanaka, Z. Phys. C67 (1995) 321.

%%% hadronic form factors in HQET
\bibitem{HFF}
M.~Neubert, \PL{B264} (1991) 455.

\bibitem{Isgur}
N.~Isgur and M.~B.~Wise, \PL{B232} (1989) 113; \PL{B237} (1990) 527.

%%% Dispersion relation
\bibitem{Caprini} 
I.~Caprini, L.~Lellouch and M.~Neubert, Nucl. Phys. B530 (1998) 153.

%%% slope parameter in BELLE
\bibitem{SPBELLE} K.~Abe {\it et.al.}, BELLE-CONF-0121 (2001).

%%% LHC
\bibitem{LHC} F.~Gianotti, talk presented at LHCC, 5 July, 2000,\\
http://gianotti.home.cern.ch/gianotti/phys\_info.html .

%%% q^2 distribution
\bibitem{Soni} K.~Kiers and A.~Soni, \PR{D56} (1997) 5786.

\end{thebibliography}
\end{document}